# Development of models for predicting Torsade de Pointes cardiac arrhythmias using perceptron neural networks

Mohsen Sharifi[1], Dan Buzatu[1]*, Stephen Harris[1] and Jon Wilkes[1]

[1]Division of Systems Biology, FDA's National Center for Toxicological Research, Jefferson, AR 72079, USA

*Corresponding author: Dan Buzatu, Ph.D.

Phone: +1 (870) 543-7304, Fax: +1 (870) 543-7287    Email: Dan.Buzatu@fda.hhs.gov

## Abstract

**Background:** Blockage of some ion channels and in particular, the hERG (human Ether-a`-go-go-Related Gene) cardiac potassium channel delays cardiac repolarization and can induce arrhythmia. In some cases it leads to a potentially life-threatening arrhythmia known as Torsade de Pointes (TdP). Therefore recognizing drugs with TdP risk is essential. Candidate drugs that are determined not to cause cardiac ion channel blockage are more likely to pass successfully through clinical phases II and III trials (and preclinical work) and not be withdrawn even later from the marketplace due to cardiotoxic effects. The objective of the present study is to develop an SAR (Structure-Activity Relationship) model that can be used as an early screen for torsadogenic (causing TdP arrhythmias) potential in drug candidates. The method is performed using descriptors comprised of atomic NMR chemical shifts ($^{13}C$ and $^{15}N$ NMR) and corresponding interatomic distances which are combined into a 3D abstract space matrix. The method is called 3D-SDAR (3-dimensional spectral data-activity relationship) and can be interrogated to identify molecular features responsible for the activity, which can in turn yield simplified hERG toxicophores. A dataset of 55 hERG potassium channel inhibitors collected from Kramer et al. consisting of 32 drugs with TdP risk and 23 with no TdP risk was used for training the 3D-SDAR model.

**Results:** An artificial neural network (ANN) with multilayer perceptron was used to define collinearities among the independent 3D-SDAR features. A composite model from 200 random iterations with 25% of the molecules in each case yielded the following figures of merit: training, 99.2 %; internal test sets, 66.7%; external (blind validation) test set, 68.4%. In the external test set, 70.3% of positive TdP drugs were correctly predicted. Moreover, toxicophores were generated from TdP drugs.

**Conclusion:** A 3D-SDAR was successfully used to build a predictive model for drug-induced torsadogenic and non-torsadogenic drugs based on 55 compounds. The model was tested in 38 external drugs.

2

# Background

Potassium plays a crucial role in the cardiovascular system. The flow of potassium across cardiomyocytes is essential for cardiac rhythm. A number of compounds block cardiac potassium ion channels and cause arrhythmia. Cardiac potassium channels have an important role in ischemic pre-conditioning. Among these channels, the ATP sensitive potassium ion channels which are ligand-gated channels can be abundantly found in all regions of the heart (Grant, 2009). Figure 1 illustrates schematic changes in voltage (cardiac membrane potential) across the cardiomyocytes. After potassium ions begin to flow into the myocytes, calcium and potassium ions offset each other, which produce a plateau phase. Further, after efflux of more potassium ions outside the myocytes, repolarization will take place. Repolarization involves interactions among numerous calcium, sodium, and potassium channels. Nevertheless, potassium channels play a key role in a type of drug-induced cardiac arrhythmia and may lead to a potentially life-threatening condition termed "Torsade de Pointes" (TdP). The lower green arrow in Figure 1, indicates a longer action potential (320 milliseconds). QT prolongation is a special cardiovascular safety concern. The QT interval characterizes the time from the depolarization to ventricular repolarization, and its elongation causes cardiac arrhythmia (Villoutreix and Taboureau, 2015).

To our knowledge, 80 voltage-gated potassium-channel families have been recognized in the human genome (Wulff et al., 2009; Szabò et al., 2010; http://vkcdb.biology.ualberta.ca/). Based on structure and function, potassium channels generally are separated into the following major categories: the voltage-gated channels with six transmembrane domains; inwardly rectifying channels with two transmembrane domains; and Tandem Pore channels with four transmembrane domains (Buckingham et al., 2005). The hERG-gene and similar variants are the most common potassium ion channels in mammals. Blockage of the hERG potassium channels can act as a trigger to cause syncope and sudden death in rare cases (Glassman and Bigger, 2001). The level of inhibition of the hERG gene is one of the earliest preclinical markers used to predict the risk of a compound causing TdP (Mirams et al, 2011).

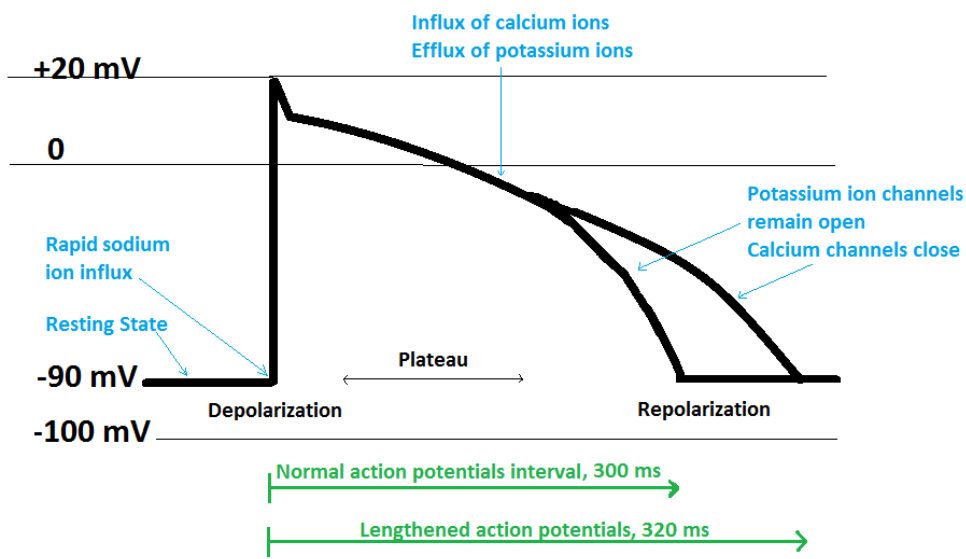

**Figure 1.** Schematic changes in ventricular action potential at the molecular level within the cardiomyocytes. The resting membrane potential of cardiomyocytes is about -90 millivolt (mV) while at the full depolarization it can be gradually shifted to +20 mV. In the repolarization stage, membrane potential will return to -90 mV. Some drugs can prolong the duration of normal action potential (lengthened action potential in green) which eventually can lead to drug-induced arrhythmia. Consequently, production of lengthened action potential (long QT syndrome) may initiate TdP arrhythmia (Adapted with permission from Sharifi, 2017).

As stated earlier, drug-induced blockade of the cardiac ion channels, especially hERG potassium channel delays cardiac repolarization and causes cardiac arrhythmia, that occasionally causes a potentially life-threatening arrhythmia (TdP), observed as "twisted points" (French "Torsade de Pointes") on the electrocardiogram (Figure 2). TdP is a particular type of atypical heart rhythm that can lead to ventricular fibrillation and sudden cardiac death. In the past, TdP was observed idiosyncratically, only after a large number of patients were exposed to a new drug. In the electrocardiogram presented on Figure 2, the patient was on therapeutic dose of methadone with a low serum potassium level of 3.1 mmol/L (normal level = 3.5-5.0 mmol/L). The adverse event (QT prolongation and in some cases TdP) was reported in a study by Pearson and Woosley describing a total of 5,503 reports of adverse events associated with methadone (43 patient noted the occurrence of TdP and 16 patients QT prolongation) (Pearson and Woosley, 2005). Methadone is metabolized in hepatocytes primarily by cytochrome P450 (CYP3A4) (Ferrari et al., 2004). A methadone derivative, levacetylmethadol, was withdrawn from the European market after being associated with TdP. Pearson and Woosley reported a case that drug-drug interactions between nelfinavir (a potent CYP3A4 inhibitor) and methadone initiated TdP (Pearson and Woosley, 2005). To date, not many studies have been conducted on drug-drug interactions between methadone and other drugs and their association with TdP arrhythmias. However this type of interaction leading to TdP is well documented for cispride and terfenadine. Both drugs are associated with prolonged ventricular repolarization, high-affinity blockade inside the hERG cavity, but rarely causing sudden death (Wysowski and Bacsanyi., 1996; Monahan et al., 1990). Methadone is also presented in the dataset used in this study (where its TdP risk was correctly predicted as positive).

More recently, detection of an important indicator of proarrhythmic liability became possible in potential drug candidates. Also, once the mechanism relating potassium channel blockage to TdP was realized, the US Food and Drug Administration (FDA) added possible or high risk for TdP as a safety criterion for new drug applications (Kerns and Di, 2008). Compounds with hERG blocking liability might fail during preclinical and costly clinical trials. Hence, understanding the molecular mechanisms involved in binding of drugs to hERG channels and drug risk identification from channel blockage is now considered essential for both pharmaceutical companies and regulatory authorities. Numerous approved drugs such as the aforementioned terfenadine (antihistamine) and cisapride (a gastroprokinetic agent) were eventually recalled due to cardiac toxicity associated with blockade of hERG channel (Yap and Camm, 2003). It is well-established that the majority of potential hERG blockers prolong QT, but the converse is not so. There are a few drugs that block hERG channels without causing TdP. Verapamil, a potent hERG channel blocker is not associated with TdP (Martin et al., 2004). Further, even though all drug-induced torsadogenic compounds have a low $IC_{50}$ (strong blockers of hERG), not all hERG blockers with strong potencies lead to TdP. For example, ranolazine is an hERG channel blocker and prolongs QT, but appears not to cause arrhythmia, due to the effects on late sodium currents (Stockbridge et al., 2013). Mirams and his colleagues, tested multiple ion channel blockage, namely, sodium calcium and potassium channels for prediction of TdP (Mirams et al., 2011). They collected 31 drugs associated with varied risks of TdP, and applied numerous pacing protocols for simulation purposes. They concluded that consideration of hERG blockade is necessary, but not sufficient, to predict torsadogenic risk (Mirams et al., 2011).

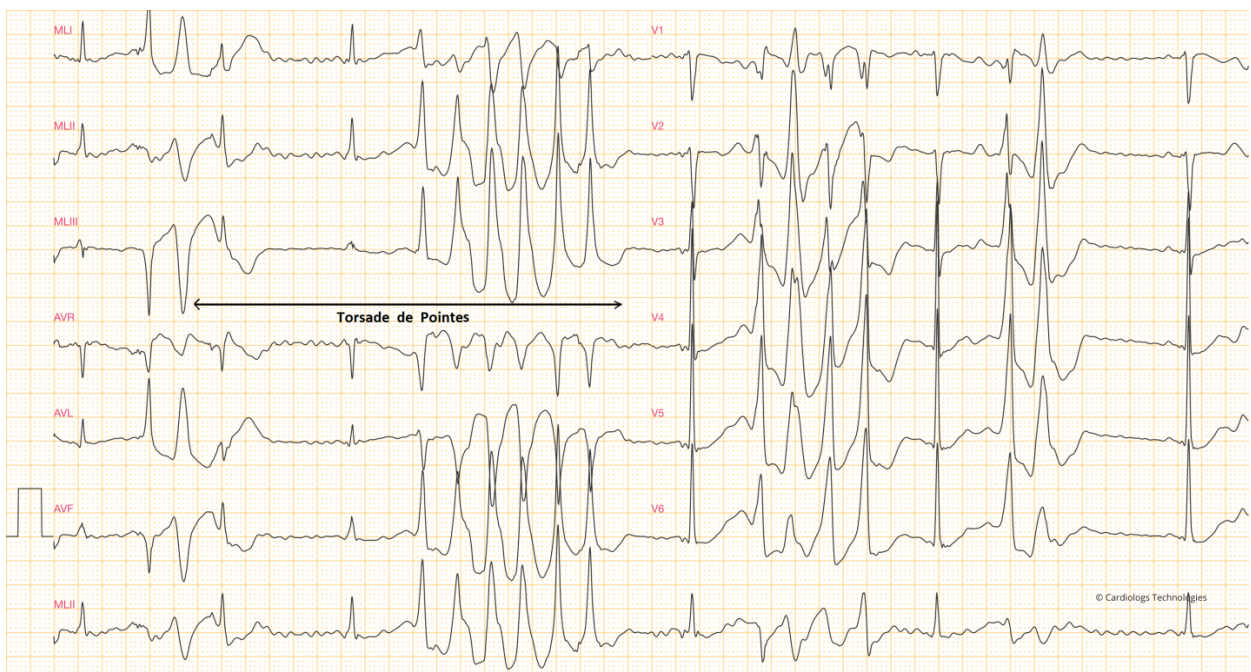

**Figure 2.** A 12-lead electrocardiogram represents a long QT syndrome in in a 49-year-old male on therapeutic dose of methadone with a serum potassium level of 3.1 mmol/L and no cardiovascular disease history seen in the patient's family. A burst of TdP can be seen on the left, and short time (under ten seconds) tachycardia can be seen on V1-V6 (right). The electrocardiogram is used with the permission of Dr. Pierre Taboulet (Taboulet, 2017).

Computational models can be used as an early screen for torsadogenic potential in drug candidates. SAR studies of hERG models as well as structure modification strategies are being developed and they aim to reduce the risk of hERG blockage. Numerous models have been built to profile potential hERG channel blockage of newly discovered compounds. Indeed, applying *in silico* tools is an emerging trend for screening and detecting potential inhibitors of hERG channels (Kerns and Di, 2008). Sanguinetti and Mitcheson studied how drugs bind as residues lining the central cavity of the hERG channel and suggested *in silico* approaches to assess hERG channel blockade (Sanguinetti and Mitcheson, 2005). Results of their models based on physicochemical properties of chemical structures used in the training set of an SAR model, indicate that electrostatic groups in the para position of phenyl rings and hydrophobic bulk on the tail of compounds both can influence drug affinity for the hERG channel (Sanguinetti and Mitcheson, 2005). In a recent study by Stoyanova-Slavova and co-workers, 237 compounds and their corresponding hERG channel activities were collected from 22 databases, and some classification models using partial least squares (PLS) regression were built where four latent variables used for reporting the results (Stoyanova-Slavova et al., 2017). The best model successfully predicted the hERG activities with an overall prediction accuracy of 0.84 for both external and internal validation sets. However, in the external (blind) test set of their models, 10 out of 16 hERG active compounds were predicted incorrectly. Further, the most important features obtained from PLS model were mapped based on PLS weights, and from these hERG pharmacophores were obtained. The diagnostically most significant endpoint for this area of modeling is TdP risk, not merely hERG binding.

The objective of the present study is to develop an SAR-like model based directly on TdP clinical adverse events, a model that can be used as an early screen for torsadogenic potential in drug candidates. The less familiar method is called 3D-SDAR and uses descriptors comprised of NMR chemical shifts ($^{13}$C and $^{15}$N) and their corresponding interatomic distances which are combined into a 3D abstract space matrix. Such models can be interrogated to identify molecular features responsible for the activity. If based on hERG blockage data, they can yield simplified toxicophores for potent inhibitors of hERG potassium cardiac ion channels. If, as in this work, the models are based directly on TdP data, they should yield toxicophores for TdP. Alternative modeling approaches included SAR and QSAR techniques embodied in commercial software packages.

## Methods

### Dataset

The 55 compounds used for training and internal test sets were obtained from the literature (Kramer et al., 2013), which include 32 torsadogenic and 23 non-torsadogenic drugs from multiple classes. The drug-induced torsadogenic risk of each drug for training and internal test sets presented in this paper (55 drugs), were originally evaluated based on the classifications assigned in Redfern et al. and the Arizona Center for Education and Research on Therapeutics (ACERT, www.azcert.org). Torsade risk results for all of the 55 compounds were from a single lab (Chan Test Corporation, Ohio). A subsampling technique was used for defining training and internal validation sets in the ANN, with 200 epochs, which means our aggregate model presented the median of predictions from the 200 individual models. Further, to find a dataset for an external test set (a "blind" validation exercise), the Essential Drug Safety Resource from PharmaPendium's advanced search engine (www.pharmapendium.com) was used and a total of 527 reports for individual compounds associated with drug-induced TdP arrhythmia were found. Further, these compounds were filtered and only compounds with more than 25 post-marketing reports (defined by the Adverse Event Reporting System (AERS)) on TdP arrhythmias were retained. For example, warfarin (anticoagulant) was reported to be associate with TdP in only five cases, which due to the low number of reports, authors assume that the TdP caused is possibly either due to other medications that patients had or because of warfarin's drug-drug interactions with other drugs that patient used concurrently. As a result of final filtering, 38 drugs (out of 527) were retained and were selected to be used in the external test set. The drugs used in this study for modeling purposes and their TdP risk are listed in Supplementary Table 1.

**Data preparation process**

For each compound in any of the data sets, a 3D mol file was downloaded directly from ChemSpider (http://www.chemspider.com/), cleaned and geometrically optimized. Energy minimization was applied using the AM1 semiempirical Hamiltonian provided by MOE software, version 2016.08 (Chemical Computing Group, Montreal, CA). Then the mol files were imported into the ACD/NMR predictor (version 12, ACD/Labs Toronto, Canada), with each molecule's atom numbering system preserved. The NMR spectra of the compounds in the dataset were generated using the HOSE algorithm (Bremser, 1978; Meiler et al., 2002). The HOSE algorithm prediction uses a 2D substructural unit. When these shifts for atom pairs in a molecule were combined with corresponding interatomic distances, the abstract pattern became 3 dimensional and in that way reflects that molecule's Cartesian 3D nature. Distances were calculated from 3D mol files using an in house program written in R and facilitated with R studio (Wilkes et al., 2016). NMR chemical shifts for atoms of interest for $^{13}$C and $^{15}$N were obtained in the Spectrus software package (ACD/Labs package, Toronto, Canada) and used as the electrostatic component of the SDAR molecular descriptors.

**Binning parameters and fingerprint construction in R**

The carbon to nitrogen bin width ratio was set to 2.5 (using 2.5 times C's bin width). Regarding bin occupancy range, for nitrogen, a shielding range between -356 and -11 PPM, width = 345 PPM and with a midpoint of -183.5 was considered. For carbon, shielding range varied from -4 to +204 PPM, width = 208 PPM and a midpoint of 100. Figures S1 and S2 in Supplementary Material illustrate carbon and nitrogen shift frequencies, respectively, for the initial 55 compounds. The NMR chemical shifts are expressed in parts per million (PPM) units with positive values for carbon and negative values for nitrogen. Supplementary Figure 1 shows the distribution of NMR chemical shifts for carbon and nitrogen. The frequency of interatomic distances between atoms in all molecular structures used in training the models is shown in Supplementary Figure 2. Further, to generate the feature matrix, descriptors were first scaled in R studio, and then binned. For statistical analysis, the data were imported into Statistica Data Miner version 11 (StatSoft, Tulsa, OK) to be modeled using its artificial neural network (ANN) algorithm. Based on the original dataset (Kramer et al., 2013), activities of drugs were assigned to "1" for drugs with risk of TdP or "0" for drugs that do not cause TdP. In the ANN, activities were input, and binned numerical values obtained from NMR chemical shifts and interatomic distances constituted descriptor vectors for each molecule.

## Selection of neural network parameters

Neural networks are non-parametric modeling tools and use a series of weights and hidden neurons to capture complex correlation between the predicted inputs and target variables. We used a Multilayer Perceptron (MLP) as multi layered feed forward neural network type with the gradient descent algorithm in the ANN. To explore parameter choices for hidden and output neurons, the following activation functions were examined: 1. Identity function (with this function, the activation level is passed on directly as the output); 2. Hyperbolic Tanh (which is a symmetric S-shaped (sigmoid) function whose output lies in the range of -1 to +1). The number of layers for the network was set to 2 and the learning rate was set to 0.1. The error functions were specified to be used in the training network and were calculated by sum of square (given by the sum of differences between the target and prediction outputs defined over the training set) using the following equation: Error = $\sum (y_i - t_i)^2$ where $y_i$ is the prediction (network outputs) of the target value $t_i$ and target values of the $i^{th}$ data point (Hill and Lewicki., 2006). In order to avoid overfitting, the "weight decay" option and advanced stopping conditions were enabled in Statistica to improve generalization performance.

## 3D Toxicophore identification for TdP Arrhythmias

Detecting the key features (so called toxicophores) associated with a biological activity entails encoding chemical structural features which can be abstracted into a 3D space matrix. Since toxicophore schemes are sensitive to the protonation state of the molecules, strong acids or bases were deprotonated. For similar reasons, we chose the "enumerate tautomers" option for the weak acids and bases. The prepared structure of mol files were ionized at neutral pH (7.0) before generating toxicophores.

*Generating Toxicophore using MOE software*
3D toxicophore generation is an essential step used in feature identification of active (drug-induced torsadogenic) drugs. To build a Toxicophore Query to match a set of torsadogenic drugs, there is an assumption that all of the molecules bind in a similar conformation to the receptor, a query that represents a toxicophore hypothesis. A Query is a collection of features, feature constraints, and volume restrictions that is applied to the annotation and atoms of a ligand conformation (Molecular Operating Environment, MOE, 2017). Firstly, mol files for all of the potential cardiac ion channel inhibitors (all drugs in the dataset causing TdP arrhythmias) were copied to MOE window, and then we ran the Flexible Alignment and obtained seven conformations where the amines and rings were overlaid. Later, using the Consensus option in the toxicophore query editor, we selected the features which matched all the molecules.

*Generating a Toxicophore in Schrödinger suite*
Similar to MOE, a 3D database that includes toxicophore information was prepared first in Maestro interface (version 10.6; https://www.schrodinger.com/maestro), and then we searched the database for matches. Then to identify features using Phase (Schrödinger's toxicophore generation module) the hydrophobic groups, hydrogen bond donors and acceptors, and aromatic rings were used as elements of a "hypothesis". Then, a common scaffold alignment was performed among the active drugs where structures were aligned on the scaffold, with conformational variation of the side chains. In this way, a hypothesis finding common toxicophores was created and scored. Finally, a toxicophore was generated from the common toxicophore hypotheses and using the top alignment scores.

*Generating a Toxicophore using a Feature (important bins) Visualizer for 3D-SDAR*
Significant bins obtained from the ANN model were projected onto 3D molecule diagrams using scripts in R studio, where NMR chemical shifts (representing electrostatic information) together with interatomic distances (steric information) were combined and tessellated into a 3D abstract matrix. In this way, visualized data was interrogated to identify molecular features responsible for the activity. Unlike the other two methods, 3D-SDAR fingerprints are invariant under rotation of the Cartesian coordinates and therefore independent of an assumed relationship between each ligand and it's hypothetical fit into a biological receptor.

## Results and Discussions

In Statistica, the output summary shows the number of hidden units each network had; test, training and validation performance (percentage of compounds predicted correctly); and test and training error values. Based on confusion matrix, average (200 iterations, hence 200 ANN models) of correctly predicted drugs in training, internal and external (blind) test sets were 99.2%, 66.7% and 68.4% respectively. In the external (blind) test set, 70.3% of positive TdP drugs (drugs that are causing TdP arrhythmias) were correctly predicted. The ratio of correctly predicted for the active drugs in the dataset is utmost important, because toxicophores are constructing and mapped from active drugs in the dataset only, therefore, to have a robust model that later can be used to construct toxicophores, the ratio of correctly predicted "active" drugs (i.e. TdP+) must always be considered.

Results of the aggregated predictions for all 200 ANN 3D-SDAR models can be seen in Supplementary Table 2, where values highlighted in red represent those which predicted incorrectly. Further we performed sensitivity analysis based on ANN results, which shows how strongly certain variables affected the particular network. In a sensitivity analysis summary, variables are sorted by sensitivities (so the first variable has the highest sensitivity and contributed most to the particular network). The significant variables later were loaded in the Feature (bins) Visualizer application as part of the process by which toxicophores were constructed.

Receiver Operating Characteristic (ROC) curves are widely used as a tool to evaluate classification models. An ROC curve represents the quality of models by visualizing the "true" positive versus the "false" positive rate. Figure 3 depicts the 200 ANN models for training, test, and validation sets. The best models show an ROC curve that approaches the left and top axes in the plot. The blue curves are so colored to indicate a large number of overlaps – the most characteristic results.

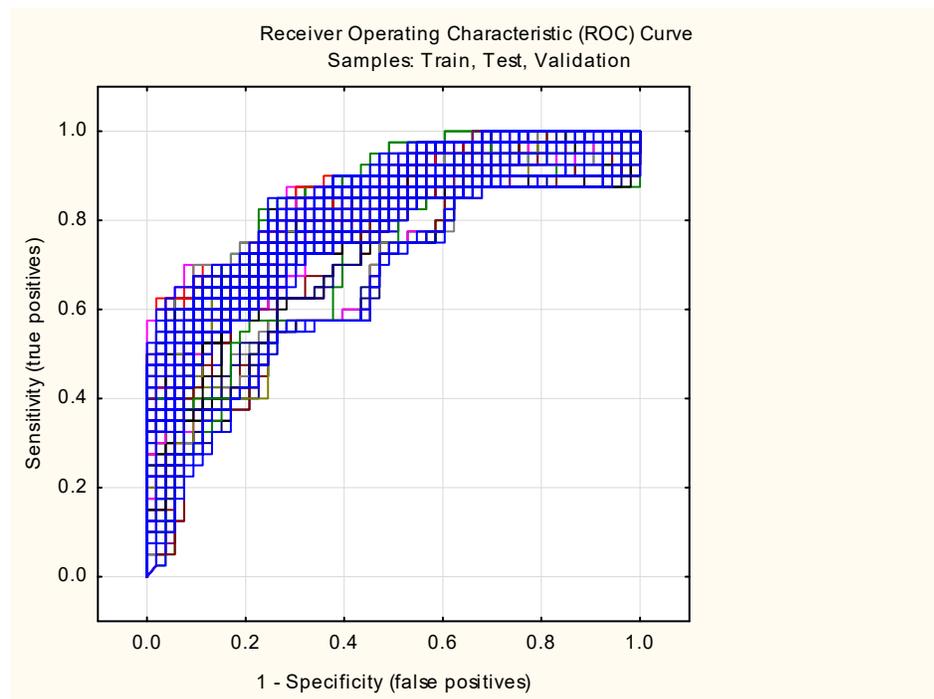

**Figure 3.** The ROC curves (cumulative) of SDAR for the training, test and validation (38 blind hold-out test drugs) sets. The majority of these models (overlapped blue lines) indicated good predictive accuracy. The lines closer to the diagonal obviously come from the predictions in the internal and external training sets, less accurate than those of the training sets.

Additionally, to have a better understanding of classification output, we classified the results on a gain plot. The gain plot shows how well a model correctly classifies a category. The larger the area between the baseline (blue) and the line for the predictive model (red) is, the better predictive accuracy can be obtained. Gain is calculated as the ratio of accurately predicted compounds to the total number of compounds. Supplementary Figures 3 and 4 show the gain charts for non-torsadogenic and drug-induced torsadogenic drugs, respectively. Comparing the two figures we see a better classification was obtained for non-torsadogenic drugs than for torsadogenic drugs. If the imbalance in the training set almost 60% torsadogenic compounds) were reflected in the predictions, one would have expected a substantial majority of torsadogenic predictions, which is not indicated in these figures. It is likely, therefore, that the toxicophores derived from the 3D-SDAR models accurately reflect the features responsible for TdP and are not simply artifacts of a statistical phenomenon, the training set imbalance.

## Mapping and Investigation of Toxicophores

In MOE, toxicophore consensus creates a list of suggested features from a set of torsadogenic-conformations of drugs. Figure 4 shows five of the most important features for active drugs (drugs with TdP risk) in the dataset.

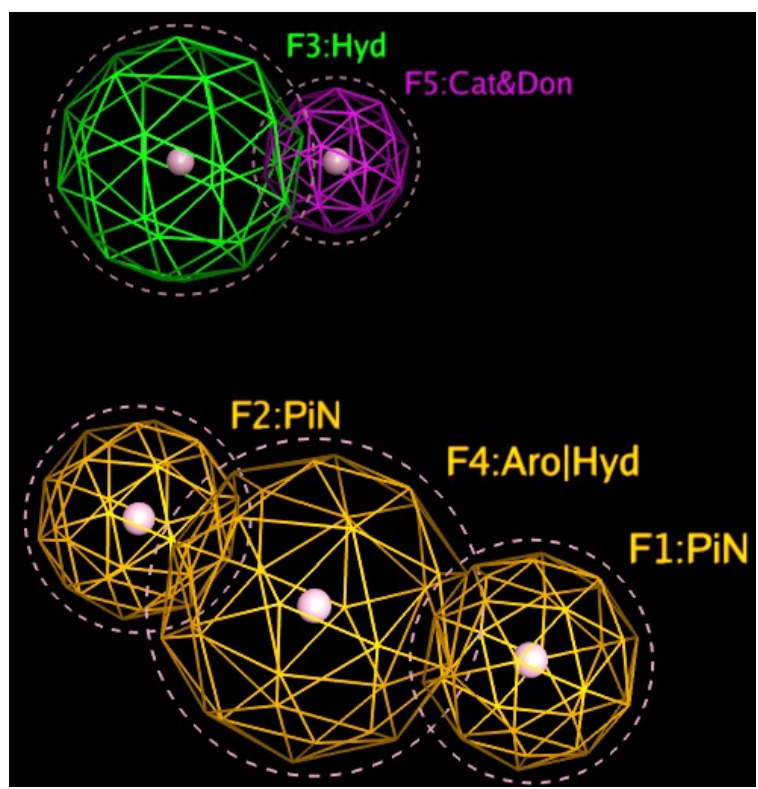

**Figure 4.** "Mesh globes" in the figure indicate the frequency with which that toxicophores appears in drugs associated with TdP risk. Each globe can be thought of as a possible toxicophore in a "3D Query panel" (MOE software). Features #1 and #2 represents two planar Pi rings. Feature #3 (green) illustrates a hydrophobic centroid (Hyd) and features #4 elucidates a Polar-Charged-Hydrophobic scheme assigns the label Aro (aromatic center) to a hydrophobic area. Feature #5 (F5) shows a cationic atom and a H-bond donor. Identification of these five features is not dependent on any of normal QSAR modeling. Therefore, the fact that these show up with some frequency in a subset of the data that includes only TdP positive compounds may or may not indicate that they are necessary participants causing TdP.

Some antipsychotic drugs such as clomipramine, nortriptyline and desipramine (tricyclic antidepressants), have a cyclopentazepine feature and all are categorized as drugs with possible

TdP risk based on ongoing systematic analysis of all available evidence stratified for AZCERT (http://www.azcert.org, as of May 2017). Since these three compounds were not included in the dataset of this study, some of the potential toxicophores (see F1, F2 and F4 in Figure 4) produced by our models could signal molecular characteristics that may indicate a problem – here likelihood of causing TdP.

An early hERG pharmacophore was introduced by Ekins and co-workers, with 15 compounds collected from the literature where four hydrophobes and a positively ionizable feature, with approximate distances between the positive center and the hydrophobes of 5-7 angstroms was proposed (Ekins et al., 2002). More recently, a catalyst hERG pharmacophore model (based on 18 publicly available hERG blockers) with a positive ionizable feature, one aromatic hydrophobic and two hydrophobic features was generated (Kratz et al., 2014).

Aligned TdP drugs with their correspondent toxicophores are illustrated in Figure 5. These mesh globes are similar to but not identical with the ones discussed above.

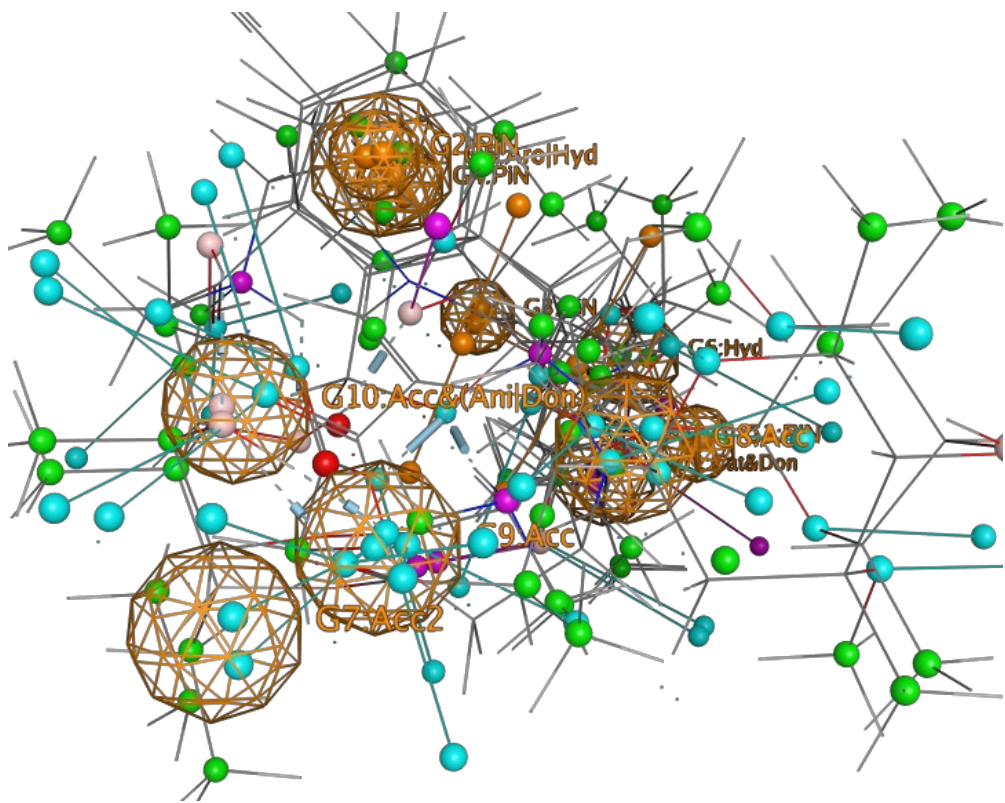

**Figure 5.** Visualization of the alignment of structures in the 59 active TdP drugs from the modeling or external validation sets studied here. The visualization comes from MOE software. The TdP positive drug molecules were aligned with respect to their potential toxicophores identified in Figure 4.

Common toxicophores obtained from Phase (Schrödinger) is shown in Figure 6 depicting three most important features on the structures (an aromatic ring in orange, an oxygen acceptor (red) and a hydrophobic atom in (green)).

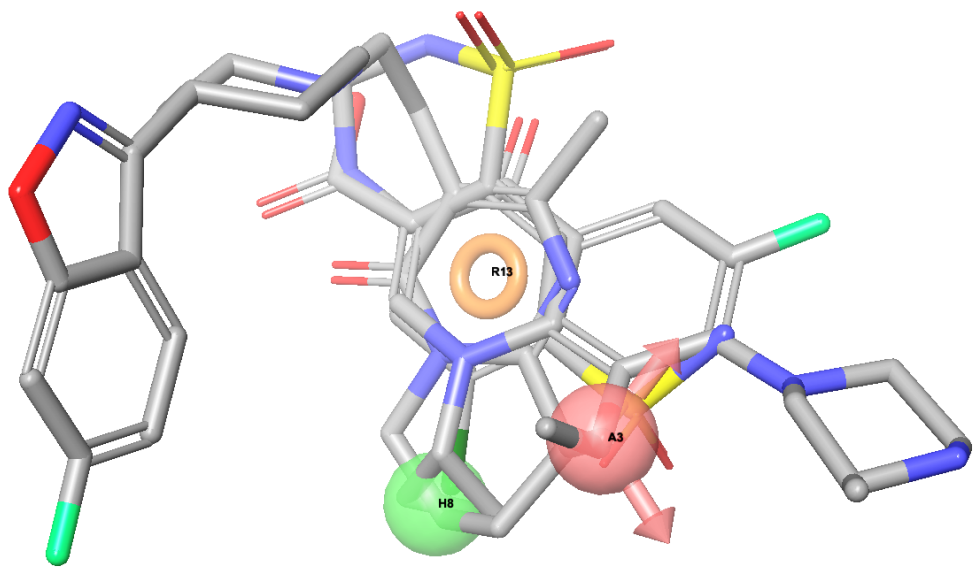

**Figure 6.** Shown is the "Common Toxicophore" from Phase (Schrödinger's toxicophore generation module) for three typical TdP positive compounds. The Common Toxicophore is generated from alignment of all available structures that are positive for causing the effect, here TdP. An aromatic ring in orange (R13), an oxygen acceptor (A3) and a hydrophobic atom (HB in green) are identified as important features. These features are selected from a list that includes the types typically important in biological activity. The selection of the important subset of features does not represent any modeling. Rather they are the features that could be associated with TdP based on the frequency with which they occur in the data set of known torsadogenic compounds.

For 3-D SDAR identification of important features might seem similar to the process described above for MOE and Schrödinger. A significant difference is that "important" SDAR bins are discovered based not only on their frequency of occurrence in the compound sets but also and primarily on the statistical weight of their contributions to the SDAR models.

Three 3D-SDAR toxicophores are constructed from the most important predictors (i.e. significant bins) and presented in Figure 7 overlaid on three representative TdP positive compounds. The toxicophores are characterized by pairs of particular elements with particular chemical shifts and the range of interatomic distances. In the case of aromatic rings, we recalculate interatomic distances so that they are shown relative to the centroid of the ring, rather than to its particular atoms. These toxicophore components reflect contributions from all modeled compounds and reflect the consensus of the 200 ANN models. They are here illustrated by their superposition as dotted grey lines above a single exemplary molecular structure. On the left feature, a benzene ring and a nitrogen atom are linked with 7-8 angstroms distance (approximately four successive carbon-atoms long, which may filled by other atoms in the structures). To the best of our knowledge, the earliest hERG pharmacophore was published in 1992 on a class III antiarrhythmic drugs (this pharmacological class typically produce TdP), and a simple pharmacophore structure of a para-substituted phenyl ring exposed to a nitrogen (through a four-atoms linking chain) was obtained from SAR model (Morgan and Sullivan, 1992; Chadwick and Goode, 2005). The middle structure in Figure 7, demonstrates a larger feature with two joint fragments (phenylmethanamine fragment is joined with a dimethylethanamine). Feature on the right is a 4-(diphenylmethyl)-piperidine toxicophore with distances between benzene rings and centroid pyridine of 4-5 angstroms. Kramer and his colleagues collected 113 compounds from the literature (where 51 compounds hit the pharmacophore) and generated several pharmacophores for the hERG model (Kramer et al., 2008). In one of the key pharmacophores obtained from their models, two aromatic hydrophobic features linked to another hydrophobic (a ring) feature was observed which is similar to the diphenylmethyl-piperidine obtained from 3D-SDAR in this paper (Figure 7).

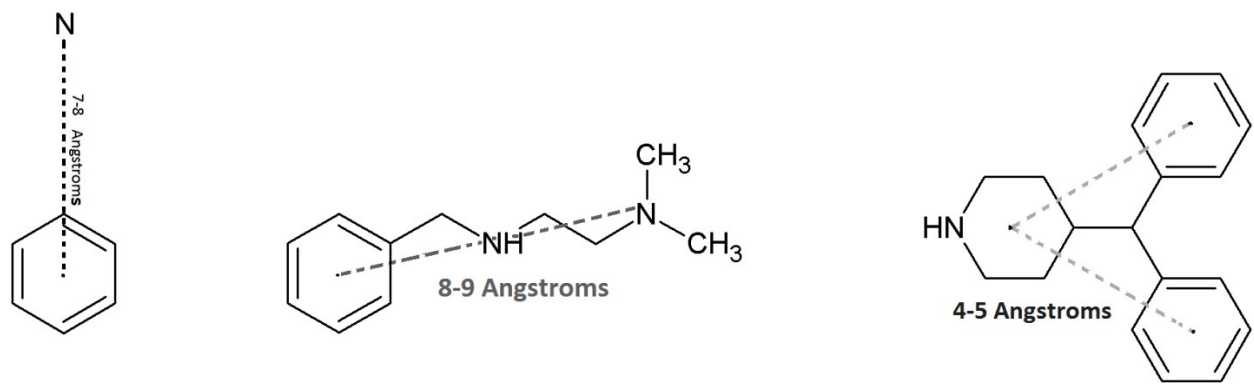

**Figure 7.** Most significant features identified by SDAR technique. The structure on the left shows a benzene ring and nitrogen atom with a distance of 7-8 angstroms (approximately equal to a sequential four-atom distance). The middle structure shows a phenylmethanamine fragment joined with a dimethylethanamine portion. On the right is 4-(diphenylmethyl)-piperidine.

As shown earlier, three different techniques were used for toxicophore mapping in this study. Toxicophores in Figures 4-6 were obtained from traditional QSAR techniques i.e. flexible alignment of TdP drugs followed by obtaining the pharmacophoric structural features (e.g.

aromatic ring, hydrophobic areas, charge interactions, etc.) for detection of active drugs, while constructed toxicophores in Figure 7 were based on the 3D-SDAR technique. Recall that the most important selected atoms (atoms of the features presented in Figure 7) are driven directly from the most significant bins of the model results, which means they represent atom data (obtained from NMR chemical shifts and their distances between pair of atoms) which shows importance of those atom pairs in the structures related to their biological activities (here TdP risk). Hence, this ability of the 3D-SDAR model is considered an advantage of 3D-SDAR approach compare to the traditional QSAR pharmacophore and toxicophore identification techniques. Moreover, unlike QSAR, 3D-SDAR technique can differentiate structural isomers (compounds with the same formula but different biological activities, e.g. cis-trans or alpha/beta isomers) because descriptors used in SDAR have different NMR chemical shift information for different structural isomers. In regards to the diverse set of toxicophore features for TdP risk from these computational models, it's worth mentioning that the hERG binding pocket is promiscuous for drug-like compounds, besides, the binding cavity volume of hERG is also large (Pearlstein et al., 2003). Another explanation resides in the different ways that the toxicophores are determined. In the first two cases (MOE and Schrödinger), they are defined only by their frequency of appearance among the available positive compounds. They are derived from characteristics of the modeling and validation sets and not from the activity models. In the SDAR case, a toxicophore meets statistical criteria of association with the compound's activity as discovered in the models and then, among the features so qualified, a frequency of occurrence filter is applied.

Nearly 83% percent of all drugs in the dataset were predicted correctly (predictions in Supplementary Table 2), and only 16 drugs (out of 93 drugs used in this study) were misclassified (predicted incorrectly). Considering a very limited sample-size used in this study (training set consisted of 43 molecules, and the internal test contained 12 drugs only), ANN performance of the 3D-SDAR model for both potential torsadogenic drugs (*i.e.*, drugs with potential TdP risk) as well as the prediction accuracy (portion correctly predicted) for non-torsadogenic drugs, is considered decent and promising.

## Conclusions

The drug-induced cardiotoxic adverse effects with risk of QT prolongation and TdP arrhythmias signify a major need in clinical studies of drug candidates (Taboureau and Jørgensen, 2011). Detection of compounds that potentially block the human hERG potassium channel is a necessary part of the drug safety process because drug-induced QT prolongation caused by these blockers occasionally appears as an adverse effect of pharmacotherapy. The models were tested using the external set of 38 drugs with TdP arrhythmias information extracted from patients' reports from the Adverse Event Reporting System (AERS)) on TdP. The torsadogenic risk of each drug for training and internal test sets was evaluated based on the classifications assigned as explained by Redfern et al., where zero "0" represents non-torsadogenic drugs and one "1" denotes drug-induced torsadogenic drugs. Decent predicted values for drugs causing TdP arrhythmias and for safe drugs (non-torsadogenic drugs) demonstrate that 3D-SDAR descriptors have predictive power, which indicates that this 3D-SDAR model contains information linking of potential inhibitors of hERG potassium channels and TdP. The TdP arrhythmias 3D-SDAR model provides comparable predictive performance to previously reported QSAR models. 3D-SDAR modeling was successfully used to build an ANN predictive model of torsadogenicity for drugs with potassium channel blocking potential. The model developed in this study can be used to evaluate TdP risk liability and also can be used in virtual screening of libraries to identify compounds with cardiac toxicophores. Toxicophores are able to capture key features and interatomic distances and can be used in filtering out compounds with elevated risk of TdP arrhythmias at an initial stage of the drug discovery process (target identification and validation) to reducing attrition in development in the pharmaceutical industry. Similarities between some of the hERG pharmacophores discovered using SDAR and published in the literature and those for TdP presented in this paper are remarkable and may point to the homologous structural functions within the molecules. On the contrary, a lack of consistency between the TdP features identified by the three different approaches, suggest that how one finds a toxicophores depends on the way one searches for it. Finally, it is not unlikely that hERG blockage and TdP may happen via different mechanisms, and therefore, to assess TdP risk, different computational tools may be required.

**Future Work and a Limitation**

In terms of hERG blocking potencies and their relevant biological activities, there are large *in vitro* datasets of inhibitor/non-inhibitor type, some of which are proprietary data and some (smaller datasets) are available in the literature. The dataset used in this paper, can be populated with more in vivo data as they become available. This involvement of cutting edge SDAR models along with classic QSAR model development, as well as drug-enzyme docking methods comprises future work in this research study. In order to further confirm the external applicability and predictive ability of these TdP risk models, additional compounds will be used in the external validation set to test the constructed models. A necessary practice in modeling intended for long term use, is to investigate diversity of the compounds in the dataset and to define the applicability domain of the models. Furthermore, it will be pertinent to ensure that datasets are robust with respect to the endpoint values contributing to the models. For instance, for TdP substrates, the goodness of the methods used for the measurement of activity should be scrutinized, and for model building, several sources of data should be compared using only compounds that have been repeatedly identified in several studies as either positive or negative for TdP. The idea behind ANNs is inspired by nature, and patterned after the human brain's network of neurons. In addition to the computational method used in this study (i.e. ANN), other modeling techniques can be used. Examples of such alternatives include PLS (supervised), PLS-DA (Partial Least Squares Discriminant Analysis), support vector machines and semi-supervised learning methods. Semi-supervised learning is a class of supervised learning techniques that makes use of unlabeled data for training and has emerged as an exciting new direction in machine learning research. It can improve models generalizability and applicability by predicting the values for unknown compounds. Other work could model not only TdP but quantitative risk of cardiac mortality or frequency of less catastrophic proarrhythmic side effects.

Among limitations, multichannel blockade of channels in addition to the potassium cardiac channel is a recognized phenomenon that would not be reflected in the hERG channel alone. Sodium and calcium channels (highly localized in the cardiac myocytes) play a key role in the electrical excitability of cardiomyocytes. Using computational models for TdP risk with biological endpoint data obtained from multichannel blockage, should generate more accurate predictive results than either qualitative TdP or hERG alone. Eventually, multichannel models may lead to better distinction between safe and unsafe drugs.

## Acknowledgements

M. Sharifi would like to thank Dr. Suguna Devi (NCTR) for help with generating Schrödinger's toxicophore, Dr. Andrew Henry (principal scientist at Chemical Computing Group) and Dr. Pierre Taboulet (cardiologist at Saint-Louis Hospital, Paris, France) for providing QT syndrome ECGs. The authors are grateful and would like to thank Dr. Donald Jensen who assisted us with this article, and likewise Drs. Vikrant Vijay and Harsh Dweep for constructive comments. M. Sharifi acknowledges support of a fellowship from the Oak Ridge Institute for Science and Education (ORISE), administered through an interagency agreement between US Department of Energy and the FDA.

## Authors' contributions

All authors conceived, designed, wrote and approved the final manuscript. All authors have contributed to the content of this paper.

## Ethics approval and consent to participate

Not applicable.

## Consent for publication

Not applicable.

## Availability of data and materials

The data are available within the manuscript and the Supplementary Tables.

## Competing and declaration of interest

## Funding

Publication of this article was funded by Division of Systems Biology, FDA's National Center for Toxicological Research, Jefferson, AR 72079, USA

Wilkes JG, Stoyanova-Slavova IB, Buzatu DA. 2016. **Alignment-independent technique for 3D QSAR analysis.** *J Comput Aided Mol Des* 2016, **30**(4):331-45.

Wulff H, Castle NA, Pardo LA. **Voltage-gated potassium channels as therapeutic targets.** *Nat Rev Drug Discov* 2009, **8**(12):982-1001.

Wysowski DK, Bacsanyi J. **Cisapride and fatal arrhythmia.** N Engl J Med 1996, 335(4):290-1.

Yap YG, Camm AJ. **Drug induced QT prolongation and torsades de pointes.** *Heart* 2003, **89**:1363-1372.

**Materials and Methods:**

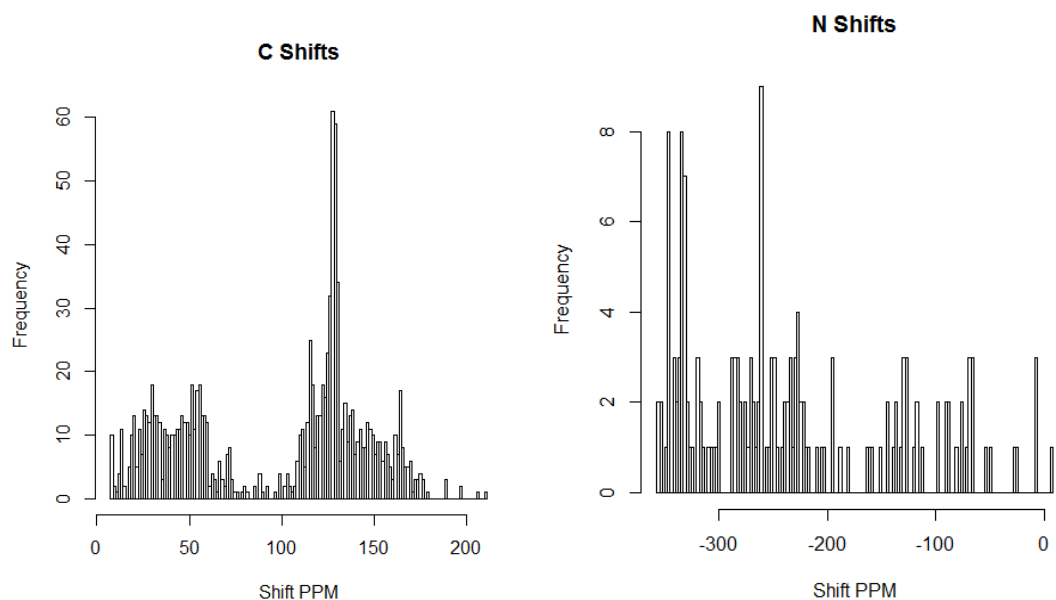

**Supplementary Figure 1.** Distribution of NMR chemical shifts for carbon (left) and nitrogen (right)

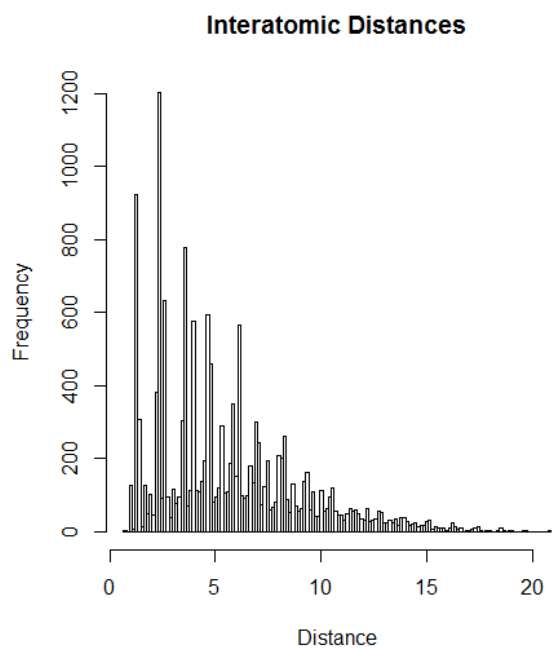

**Supplementary Figure 2.** Distribution of the interatomic distances between pairs of atoms, which play a key role in the formation of SDAR features.

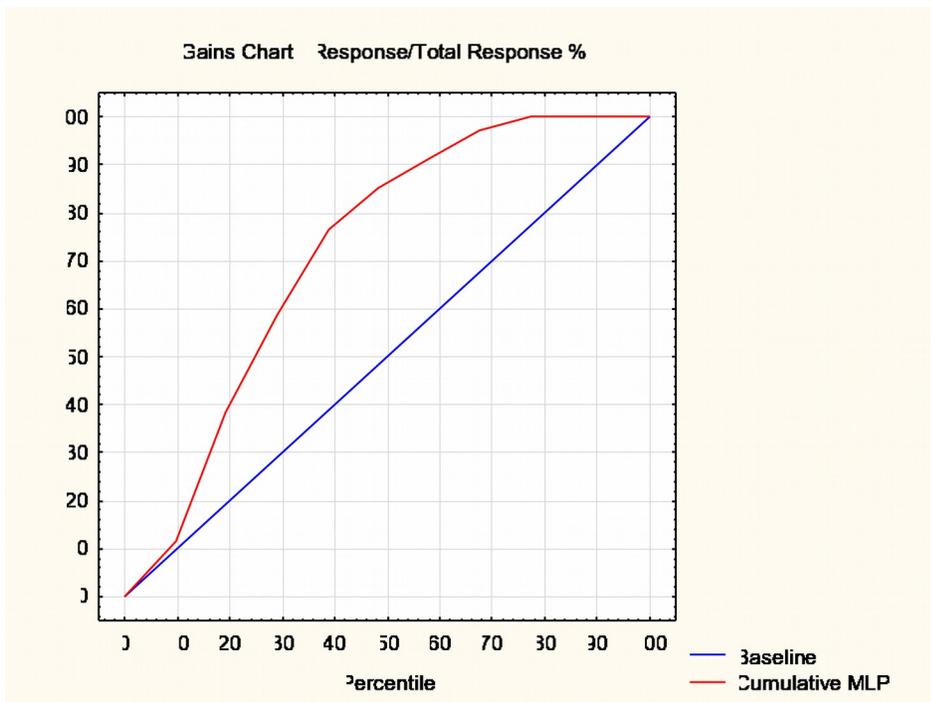

**Supplementary Figure 3.** Gain chart for non-torsadogenic drugs

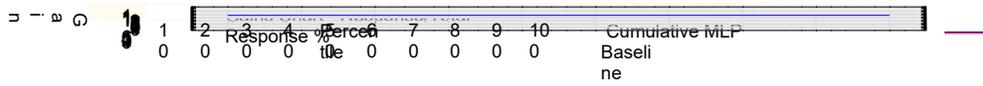

**Supplementary Figure 4.** Gain chart for torsadogenic drugs

24